\documentclass[lettersize,journal]{IEEEtran}
\usepackage{amsmath,amsfonts}
\usepackage{algorithmic}
\usepackage{algorithm}
\usepackage{array}
\usepackage[caption=false,font=normalsize,labelfont=sf,textfont=sf]{subfig}
\usepackage{textcomp}
\usepackage{stfloats}
\usepackage{url}
\usepackage{verbatim}
\usepackage{graphicx}
\usepackage{cite}
\usepackage{hyperref}

\usepackage{multirow}
\usepackage{booktabs}
\usepackage{xcolor}

\begin{document}

\title{Comparing the Pretrained Models of Source Code by Re-pretraining Under a Unified Setup}

\author{Changan Niu, Chuanyi Li, Vincent Ng, Bin Luo
    \IEEEcompsocitemizethanks{
        \IEEEcompsocthanksitem C. Niu, C. Li and B. Luo are with the State Key Laboratory for Novel Software Technology at Nanjing University, Software Institute, Nanjing University, 22 Hankou Road, Nanjing, China.\protect\\
        E-mails: niu.ca@outlook.com, \{lcy,luobin\}@nju.edu.cn.
        \IEEEcompsocthanksitem V. Ng is with the Human Language Technology Research Institute, University of Texas at Dallas, Richardson, TX 75083-0688.\protect\\
        E-mail: vince@hlt.utdallas.edu.
        \IEEEcompsocthanksitem Corresponding Author: Chuanyi Li.
    }
    \thanks{Manuscript received XXX XX, 202X; revised XXX XX, 202X.}
}

\markboth{IEEE Transactions on Neural Networks and Learning Systems}%
{Niu \MakeLowercase{\textit{et al.}}: Comparing the Pre-trained Models of Source Code by Re-pre-training under a Unified Setup}


\maketitle

\begin{abstract}
Recent years have seen the successful application of large pre-trained models of source code (CodePTMs) to code representation learning, which have taken the field of software engineering (SE) from task-specific solutions to task-agnostic generic models. By the remarkable results, CodePTMs are seen as a promising direction in both academia and industry. While a number of CodePTMs have been proposed, they are often not directly comparable because they differ in experimental setups such as pre-training dataset, model size, evaluation tasks and datasets, etc. In this paper, we first review the experimental setup used in previous work and propose a standardized setup to facilitate fair comparisons among CodePTMs to explore the impacts of their pre-training tasks. Then, under the standardized setup, we re-pre-train CodePTMs using the same model architecture, input modalities, and pretraining tasks, etc., as they declared and fine-tune each model on each evaluation SE task for evaluating. Finally, we present the experimental results and make a comprehensive discussion on the relative strength and weakness of different pre-training tasks w.r.t. each SE task. We hope our view can inspire and advance the future study of more powerful CodePTMs. 
\end{abstract}

\begin{IEEEkeywords}
Pre-training task, supervised learning, source code
\end{IEEEkeywords}

\section{Introduction}

\IEEEPARstart{S}{upservised} learning used to dominate the field of software engineering (SE) for a long time, however, it faces many critical challenges. One of them is the insufficient of large-scale labeled data to train millions or even billions of parameters in deep neural networks. In the same way as annotating text in natural language processing (NLP), labeling SE data, such as code, can be equally costly, if not more so. Faced with the same problem, the NLP community has pioneered the use of \textit{pre-training} to solve this problem. Pre-training enables the model to first be trained on a large, easy-to-obtain unlabeled dataset by using one or more \textit{self-supervised} tasks (i.e., tasks which extract labels from the input as the supervision) to obtain linguistic and commonsense knowledge about the language. Then the pre-trained model is fine-tuned on the target task using task-specific labeled data in the standard supervised manner.

Pre-training has brought the field of NLP into a new era by enabling models to learn task-agnostic general knowledge on large-scale data, allowing a large number of pre-trained models (PTMs) to be developed and widely adopted, such as BERT~\cite{devlin2019bert}, GPT-2~\cite{radford2019gpt2}, ELECTRA~\cite{clark2019electra}, T5~\cite{raffel2020t5} and BART~\cite{lewis2020bart}. Since both code and natural language (NL) text can be viewed as a sequence of (code/word) tokens, it is also feasible to apply these PTMs to SE applications directly. However, there exist many source code-specific characteristics that prevent them from being optimal for SE tasks, e.g., unlike natural language text which consists only of words, codes consist of pure code written in a programming language (PL) and comments written in natural language, as well as syntactic structures (as defined in Abstract Syntax Trees (ASTs)) and semantic structures (as defined in Control Flow Graphs (CFGs)). Therefore, a number of pre-trained models of source code (CodePTMs) have emerged in the SE community to better take these code-specific characteristics into account.

Although the existing publicly downloadable code pre-training models are different from various perspectives such as model architectures, pre-training tasks, and pre-training setups (including pre-training datasets, model size, vocabulary size, etc.) ~\cite{codeptmsurvey}, the core innovations or essential differences of code pre-training models lie in pretraining tasks only instead of model architecture and pre-training setups. So, with the recent development/growing of CodePTMs, it is necessary to study the impact of different pre-training tasks on the performance of source code pre-training to advance the future study of more powerful CodePTMs. Since current CodePTMs offer only four distinct architectures, it is beneficial to utilize models based on the same architecture to explore the impact of pre-training tasks on performance while avoiding the interference of different architectures. However, to avoid the influence of pre-training settings, we must re-pre-train and re-evaluate current CodePTMs. The reason is four-fold.

First of all, the pre-training datasets used by current CodePTMs vary, which may include distinct data scale, distribution, PLs, etc.  PLBART~\cite{ahmad2021plbart}, for example, is pre-trained on 230M functions across Java and Python, while CodeT5~\cite{wang2021codet5}, on 8.35M functions across Java, Python, JavaScript, PHP, Go, Ruby, C, and C\#.

\begin{table*}[htbp]
    \centering
    \caption{Details of evaluation SE tasks and corresponding datasets, metrics, etc.}
    \label{table:se_tasks}
    \resizebox{\linewidth}{!}{%
        \begin{tabular}{lllllllll}
            \toprule
                \textbf{Type} &
                \textbf{I-O} &
                \textbf{Task} &
                \textbf{Ab.} &
                \textbf{Definition} &
                \textbf{Dataset} &
                \textbf{Subset} &
                \textbf{Size} &
                \textbf{Metrics} \\
            \midrule
                \multirow{5}{*}{Und.} &
                \multirow{2}{*}{C-V} &
                Defect Detection &
                DD &
                \begin{tabular}[c]{@{}l@{}}Detect if the given function contains\\ any defect.\end{tabular} &
                \multicolumn{2}{l}{Devign~\cite{zhou2019devign}} &
                21K/2.7K/2.7K &
                Acc \\
            \cline{3-9}
                &
                &
                Clone Detection&
                CD &
                \begin{tabular}[c]{@{}l@{}}Identify if two code snippets are\\ semantically equivalent.\end{tabular} &
                \multicolumn{2}{l}{BigCloneBench~\cite{svajlenko2014bigclonebench}} &
                900K/416K/416K &
                F1/P/R \\
            \cline{2-9}
                &
                C-C &
                \begin{tabular}[c]{@{}l@{}}Code-to-Code\\ Retrieval\end{tabular} &
                CR &
                \begin{tabular}[c]{@{}l@{}}Retrieve semantically similar code for\\ the given query code.\end{tabular} &
                \multicolumn{2}{l}{POJ-104~\cite{mou2016poj104}} &
                32K/8K/12K &
                MAP \\
            \cline{2-9}
                &
                \multirow{2}{*}{NL-C} &
                Code Search &
                CS &
                \begin{tabular}[c]{@{}l@{}}Find the most relevant code for a given\\ NL query from a set of candidates.\end{tabular} &
                \multicolumn{2}{l}{AdvTest~\cite{lu2021codexglue}} &
                252K/9.6K/19K &
                MRR@R \\
            \cline{3-9}
                &
                &
                \begin{tabular}[c]{@{}l@{}}Code Question\\ Answering\end{tabular} &
                QA &
                \begin{tabular}[c]{@{}l@{}}Predict whether the given code an-\\swers the given NL query.\end{tabular} &
                \multicolumn{2}{l}{\begin{tabular}[c]{@{}l@{}}CoSQA~\cite{huang2021cosqa}, WebQueryTest~\cite{lu2021codexglue}\end{tabular}} &
                20K/604/1K &
                Acc \\
            \hline
                \multirow{7}{*}{Gen.} &
                \multirow{5}{*}{C-C} &
                \multirow{2}{*}{Code Translation} &
                \multirow{2}{*}{CT} &
                \multirow{2}{*}{\begin{tabular}[c]{@{}l@{}}Translate the code in one PL to the\\ code in another one.\end{tabular}} &
                \multirow{2}{*}{CodeTrans~\cite{lu2021codexglue}} &
                Java$\rightarrow$C\# &
                10K/100/1K &
                EM/B./C.B. \\
            \cline{7-9}
                &
                &
                &
                &
                &
                &
                C\#$\rightarrow$Java &
                10K/100/1K &
                EM/B./C.B. \\
            \cline{3-9}
                &
                &
                \multirow{2}{*}{Bug Fixing} &
                \multirow{2}{*}{BF} &
                \multirow{2}{*}{\begin{tabular}[c]{@{}l@{}}Repair buggy code by generating the\\ fixed ones.\end{tabular}} &
                \multirow{2}{*}{BFP~\cite{tufano2019bfp}} &
                small &
                47K/5.8K/5.8K &
                EM/B./C.B. \\
            \cline{7-9}
                &
                &
                &
                &
                &
                &
                medium &
                52K/6.5K/6.5K &
                EM/B./C.B. \\
            \cline{3-9}
                &
                &
                Code Completion &
                CC &
                \begin{tabular}[c]{@{}l@{}}Predict the following/missing token\\ of a given code context.\end{tabular} &
                \multicolumn{2}{l}{PY150~\cite{raychev2016py150}} &
                100K/5K/50K &
                EM \\
            \cline{2-9}
                &
                C-NL &
                Code Summarization &
                SM &
                \begin{tabular}[c]{@{}l@{}}Generate a text description of the\\ functionality of a function.\end{tabular} &
                \multicolumn{3}{l}{CodeSearchNet (Filtered)~\cite{lu2021codexglue}~(see Table~\ref{table:csn_statistics} for details)} &
                B. \\
            \cline{2-9}
                &
                NL-C &
                Code Generation &
                CG &
                Generate code given NL description. &
                \multicolumn{2}{l}{CONCODE~\cite{iyer2018concode}} &
                100K/2K/2K &
                EM/B./C.B. \\
            \bottomrule
        \end{tabular}%
    }
\end{table*}

Second, although Transformer~\cite{vaswani2017transformer} is widely used by CodePTMs as the base architecture, they are differ in terms of model size, such as the model dimensions and number of layers. For example, using Transformer encoder as the architecture, CuBERT~\cite{kanade2020cubert} has a model dimension of 1024 with 24 layers and 16 self-attention heads, while CodeBERT~\cite{feng2020codebert} has a dimension of 768 with 12 layers and 12 self-attention heads. 

Moreover, current CodePTMs are typically evaluated on different downstream tasks and datasets of SE, leading us not to know their relative strengths and weaknesses. Although Lu et al.~\cite{lu2021codexglue} proposed CodeXGLUE, a benchmark for code-related tasks that contains 14 datasets across 10 tasks, no CodePTM has been evaluated on all datasets so far. In addition, although there are a number of CodePTMs that select tasks from CodeXGLUE for evaluation, some of these tasks are used by many CodePTMs, while others are left unattended. The lack of an adequate comparison on the same task and comprehensive comparisons across different types of tasks prevents us from fully knowing the relative performance among CodePTMs. 

Last but not least, there are many other experimental setup that may have an impact on the results which are also uncertainties of comparing CodePTMs to explore the impacts of pre-training tasks, such as vocabulary size, runtime settings (e.g., batch size, number of iterations, early stopping patience), input and output length limits, etc. 

Similar to that GPT-3~\cite{brown2020language} has demonstrated performance improvements on NLP tasks through the utilization of more data, larger models, and extended pre-training time, the same approach can enable CodePTMs to achieve advancements in Software Engineering (SE) tasks. However, as general researchers, we are more interested in another way to improve the performance of CodePTMs on SE tasks, which is the research question of our current work, that is, how to design more powerful CodePTMs with limited resources such as data and computation? So, one of our goals is to raise the awareness of AI audience on the impact that AI technologies -- in this case the development and use of pre-trained models -- have on SE, and important AI application domain, specifically by providing them with a comprehensive comparison of the recent development of CodePTMs.

To this end, in this paper, we propose to make an empirical comparison of existing pre-training techniques of source code by re-pre-training current CodePTMs (according to their declared model architecture and pre-training tasks) and evaluating them under a unified experimental setups including pre-training dataset, model size, downstream tasks and datasets, evaluation metrics, etc. We first introduce evaluation SE tasks and then present and briefly discuss the selected CodePTMs for re-pre-training. Next, we describe our proposed unified experimental setup and proceed to show the experimental results. Please note that we do not perform hyper-parameter search when re-pre-training, and the model is re-pre-train with a single training budget, both can have effect on the results. Based on the observed results, we conduct a more in-depth, comprehensive, and constructive discussion on the impact of pre-training tasks on the performance of pre-training source code w.r.t. each SE task. Finally, provide our outlook in this domain. We hope that this paper will provide subsequent researchers with a comprehensive and comparable insight into the current techniques present in CodePTMs, so as to facilitate their design of more powerful CodePTMs. We also have made our code publicly available to facilitate future research\footnote{\url{https://github.com/NougatCA/CodePTM\_re\_train}}.

\section{Evaluation SE Tasks, Datasets and Metrics}
\label{section:setasks}

Niu et al.~\cite{codeptmsurvey} conclude that there are totally 18 key SE tasks to which currently published pre-trained models of source code have been applied. However, some of these tasks are of the same type or are just used for evaluating one single CodePTM. So, in this paper, as is shown in Table~\ref{table:se_tasks}, we select 10 out of those 18 tasks with corresponding datasets for evaluating re-pre-trained models according to the following rules: (1) the tasks cover a variety of task types, (2) a representative and commonly used task is selected for the same task type, and (3) the dataset also needs to be widely adopted. 

As can be seen in the first two columns of Table~\ref{table:se_tasks}, we classify each task along two dimensions: (1) whether the task concerns \textit{understanding} (\textbf{Und.}) or \textit{generation} (\textbf{Gen.}); and (2) the type of input assumed by the task and the type of produced output (\textbf{I-O}), where \textbf{C}, \textbf{NL}, and \textbf{V} denote code, natural language, and extracted/predicted value, respectively.

\begin{table*}[htbp]
	\centering
    \caption{Details of CodePTMs and their reported performance on SE tasks in Table~\ref{table:se_tasks}. The inputs can be handled via a Together (\&) or Standalone (+) strategy, and the pre-training tasks can be learned jointly (\&), sequentially (;), or alternately (/).}
	\label{table:codeptms}
	\resizebox{\textwidth}{!}{%
	\begin{tabular}{llllcc|c|cc|ccc|c|c}
	    \toprule
            \multirow{3}{*}{\textbf{CodePTM}} &
            \multirow{3}{*}{\textbf{Arch.}} &
            \multirow{3}{*}{\textbf{Modality}} &
            \multirow{3}{*}{\textbf{Pre-training tasks}} &
            \multicolumn{5}{c|}{\textbf{Code Understanding Tasks}} &
            \multicolumn{5}{c}{\textbf{Code Generation Tasks}} \\
        \cline{5-14}
            &
            &
            &
            &
            \multicolumn{2}{c|}{\textbf{C-V}} &
            \multicolumn{1}{c|}{\textbf{C-C}} &
            \multicolumn{2}{c|}{\textbf{NL-C}} &
            \multicolumn{3}{c|}{\textbf{C-C}} &
            \multicolumn{1}{c|}{\textbf{C-NL}} &
            \multicolumn{1}{c}{\textbf{NL-C}} \\
        \cline{5-14}
            &
            &
            &
            &
            \textbf{DD} &
            \textbf{CD} &
            \textbf{CR} &
            \textbf{CS} &
            \textbf{QA} &
            \textbf{CT} &
            \textbf{BF} &
            \textbf{CC} &
            \multicolumn{1}{c|}{\textbf{SM}} &
            \textbf{CG} \\
        \midrule
        C-BERT~\cite{buratti2020c-bert} &
            TE &
            Code &
            MLM &
            57.4 &
            &
            &
            &
            &
            &
            &
            &
            &
            \\
        \hline
            CuBERT~\cite{kanade2020cubert} &
            TE &
            Code &
            MLM\&NSP &
            &
            &
            79.12 &
            &
            98.09 &
            &
            &
            &
            &
            \\
        \hline
            CugLM~\cite{liu2020cuglm} &
            TE &
            Code &
            IMLM\&NSP\&FLM &
            &
            &
            &
            &
            &
            &
            &
            &
            &
            \\
        \hline
            CodeBERT~\cite{feng2020codebert} &
            TE &
            Code\&Doc &
            MLM;RTD &
            &
            &
            &
            &
            &
            &
            &
            &
            17.83 &
            \\
        \hline
            GraphCodeBERT~\cite{guo2021graphcodebert} &
            TE &
            Code\&Doc\&DFG &
            MLM\&EP\&NA &
            &
            97.1 &
            &
            &
            &
            59.1 &
            13.2 &
            &
            &
            \\
        \hline \hline
            CodeGPT-adapted~\cite{lu2021codexglue} &
            TD &
            Code &
            FLM & 
            &
            &
            &
            &
            &
            &
            &
            75.11 &
            &
            20.1
            \\
        \hline \hline
            DeepDebug~\cite{drain2021deepdebug} &
            TF &
            Code &
            SMLM &
            &
            &
            &
            &
            &
            &
            15.05 &
            &
            &
            \\
        \hline
            MLM+DOBF~\cite{roziere2021dobf} &
            TF &
            Code &
            MLM;SIMLM &
            &
            95.9 &
            &
            38.3 &
            &
            &
            &
            &
            18.65 &
            \\
        \hline
            PLBART~\cite{ahmad2021plbart} &
            TF &
            Code+Post &
            DAE &
            63.18 &
            \textbf{97.2} &
            &
            &
            &
            \textbf{64.8} &
            14.10 &
            &
            18.32 &
            18.8 \\
        \hline
            T5-learning~\cite{mastropaolo2021t5-learning} &
            TF &
            Code+Doc &
            SMLM &
            &
            &
            &
            &
            &
            &
            6.5 &
            &
            &
            \\
        \hline
            ProphetNet-Code~\cite{qi2021prophetnet-code} &
            TF &
            Code+Doc &
            FNP &
            &
            &
            &
            &
            &
            &
            &
            &
            18.54 &
            \\
        \hline
            CoTexT~\cite{phan2021cotext} &
            TF &
            Code\&Doc &
            SMLM &
            \textbf{65.99} &
            &
            &
            &
            &
            &
            17.07 &
            &
            18.55 &
            20.1 \\
        \hline
            CodeT5~\cite{wang2021codet5} &
            TF &
            Code\&Doc &
            SMLM/IT/SIMLM/BDG &
            65.02 &
            \textbf{97.2} &
            &
            &
            &
            63.7 &
            \textbf{18.39} &
            &
            \textbf{19.77} &
            \textbf{21.15} \\
        \hline
            TreeBERT~\cite{jiang2021treebert} &
            TF &
            Code\&AST &
            TMLM\&NOP &
            &
            &
            &
            &
            &
            &
            &
            &
            &
            \\
        \hline
            SPT-Code~\cite{niu2022sptcode} &
            TF & 
            Code\&Names\&AST &
            CAP;MASS;MNG &
            &
            &
            &
            &
            &
            62.18 &
            14.2 &
            &
            15.0 &
            \\
        \bottomrule
    \end{tabular}%
    }
\end{table*}

Besides, Table~\ref{table:se_tasks} also shows the abbreviation (Ab.), definition, dataset, subset (if any), size of the training/development/test set, and main evaluation metrics for each task. These metrics are fairly standard. For retrieval and classification tasks, metrics such as Acc (Accuracy), i.e., the ratio of correct predictions, F1, Precision (P), Recall (R), Mean Reciprocal Rank (MRR) and Mean Average Precision (MAP) are typically used. For generation tasks, metrics developed in the NLP community for summarization and translation tasks, such as BLEU (B.)~\cite{papineni2002bleu} and ROUGE-L (R.L)~\cite{lin2002rouge}, as well as variants developed in the SE community, such as CodeBLEU (C.B.)~\cite{ren2020codebleu}, are used. Moreover, the generation task also has the use of variants of accuracy as the evaluation metrics, one of which indicates that the sequence generated by the model needs to exactly match (EM) the correct answer~\cite{roziere2020transcoder}.

\section{Selected CodePTMs for Re-pre-training}
\label{section:codeptms}

Table~\ref{table:codeptms} lists all the CodePTMs to be re-pre-trained under the unified experimental setup, as well as their reported performance on SE tasks and datasets shown in Table~\ref{table:se_tasks}. These CodePTMs are classified along three dimensions, as described below.

\begin{table*}[t!]
    \centering
    \caption{Categorization and description of the pre-training tasks mentioned in Table~\ref{table:codeptms}}
    \label{table:pretrainingtask}
    \resizebox{\textwidth}{!}{
    \begin{tabular}{l|l|l|ll}
        \toprule
            \multicolumn{2}{c|}{\textbf{Type}} & \textbf{O.} & \textbf{Task} & \textbf{Full Name and Description} \\
        \midrule
            \multicolumn{2}{c|}{\multirow{8}{*}{NLA/CA}}  & \multirow{6}{*}{G} & MLM~\cite{devlin2019bert}       
            & Masked Language Modeling: predicts a certain percentage of tokens that have been randomly masked in the input. \\ \cline{4-5}
            \multicolumn{2}{c|}{}    &  & FLM~\cite{liu2020cuglm}       
            & Forward LM: maximizes the conditional probabilities of all the words by taking their previous words as contexts. \\ \cline{4-5}           
            \multicolumn{2}{c|}{}    &  & FNP~\cite{qi2021prophetnet-code}       
            & Future N-gram Prediction: a variant of FLM that involves predicting the next $n$ ($n>1$) tokens simultaneously instead of one token. \\ \cline{4-5}           
           \multicolumn{2}{c|}{}     &  & MASS ~\cite{niu2022sptcode}      
            & MAsked Seq2Seq: reconstructs the sentence fragment given the remaining part of the sentence in the encoder-decoder framework.\\ \cline{4-5}
            \multicolumn{2}{c|}{}    &  & SMLM ~\cite{mastropaolo2021t5-learning}      
            & Seq2Seq MLM: randomly masks a set of token spans in the input and sequentially predicts them in the encoder-decoder framework.\\ \cline{4-5}
            \multicolumn{2}{c|}{}    &  & DAE  ~\cite{ahmad2021plbart}       
            & Denoising Auto-Encoding: corrupts the input (by masking, deleting tokens, etc.) and uses the model to recover the original input. \\ \cline{3-5}
            \multicolumn{2}{c|}{}    & \multirow{2}{*}{C} & NSP ~\cite{kanade2020cubert}     
            & Next Sentence Prediction: determines whether two given sentences (i.e., logical lines of code) are coherent.\\ \cline{4-5}
            \multicolumn{2}{c|}{}     &  & RTD ~\cite{feng2020codebert}   
            & Replaced Token Detection: identifies the replaced tokens in the input (i.e., tokens produced by a small generator network). \\ \cline{1-5}
            
            \multicolumn{2}{c|}{\multirow{3}{*}{CA Only}}  & \multirow{2}{*}{G} & IMLM ~\cite{liu2020cuglm}      
            & Identifier MLM: an adaptation of MLM to source code that masks only the identifiers in the code text.  \\ \cline{4-5}
            \multicolumn{2}{c|}{}                     &  & SIMLM  ~\cite{wang2021codet5}  
            & Seq2Seq IMLM: an adaptation of Seq2Seq MLM to source code that masks only the identifiers in the code text.\\ \cline{3-5}
            \multicolumn{2}{c|}{}                     & C & IT  ~\cite{wang2021codet5}  
            & Identifier Tagging: determines if the input token at each position is an identifier or not via binary classification.\\ \cline{1-5}
            \multicolumn{2}{c|}{\multirow{2}{*}{SA Only}}  & \multirow{2}{*}{C} & EP ~\cite{guo2021graphcodebert}     
            & Edge Prediction: masks the edges connecting randomly selected nodes in a DFG, then predicts the masked edges.  \\ \cline{4-5}
            \multicolumn{2}{c|}{}                     &  & NOP ~\cite{jiang2021treebert}      
            & Node Order Prediction: randomly changes the order of some nodes in an AST, then determines if a change occurs.\\ \cline{1-5}
            \multirow{5}{*}{CMA} &\multirow{2}{*}{CN}  & \multirow{2}{*}{G} & BDG ~\cite{wang2021codet5} 
            &Bimodal Dual Generation: generates a NL summary if code is given, and generates code if NL is given. \\ \cline{4-5}
                                   &                       &  & MNG ~\cite{niu2022sptcode}   
            &Method Name Generation: generates the sub-token sequence of the method name based on a given method body. \\ \cline{2-5}
                                   &\multirow{3}{*}{CS}  & G & TMLM ~\cite{jiang2021treebert} 
            &Tree MLM: masks some terminal nodes/identifiers in ASTs/code on encoder/decoder side, then generates complete code sequence.\\ \cline{3-5}
                                   &                      & \multirow{2}{*}{C} & NA ~\cite{guo2021graphcodebert} 
            &Node Alignment: samples nodes in a DFG, masks the edge connecting each node to its code token, then predicts the masked edges. \\ \cline{4-5}
                                   &                      &  & CAP ~\cite{niu2022sptcode} 
            & Code-AST Prediction: determines whether the given code and AST correspond to each other.\\ 
        \bottomrule
    \end{tabular}
    }
\end{table*}

\subsection{Architecture}
\label{section:codeptm_overview_architecture}

The selected CodePTMs belong to three model architectures, namely, \textbf{Transformer} (TF)~\cite{vaswani2017transformer}, \textbf{Transformer-Encoder} (TE, i.e., corresponds to the architecture of the encoder part of TF), and \textbf{Transformer-Decoder} (TD, i.e., corresponds to the architecture of the decoder part of TF). However, there are also two CodePTMs based on \textbf{Long Short-Term Memory} (LSTM)~\cite{hochreiter1997lstm} and they are SCELMo~\cite{karampatsis2020scelmo} and CodeDisen~\cite{zhang2021codedisen}. The reason we excluded models based on the LSTM architecture from the comparison is that we cannot control their model sizes to be consistent with the model size of those based on the Transformer architecture. In our experiments, we ensure the pre-trained TE and TD models of the same size by setting them the unified input and output length limits, feed forward dimensions, number of layers and heads. For TF models, we ensure that their encoders and decoders are of the same size as TE and TD respectively. So, all TFs are naturally twice the model size of TEs/TDs. Therefore, we would only compare the TF models with TF models, or compare the TE models with TE/TD models, to explore the impact of pre-training tasks on model performance under the same model size. As a future direction, it would be interesting to evaluate the effect of matching the number of parameters in encoder-only models and encoder-decoder models, e.g., by considering wider or deeper models.

\subsection{Modality}
\label{section:codeptm_overview_modality}

CodePTMs listed in Table~\ref{table:codeptms} cover all the three input modalities utilized by current CodePTMs, i.e., \textbf{code}, \textbf{natural language}, and \textbf{structure}. For the code modality, all models utilize the sequence of code tokens. However, there is one of the current CodePTMs using the intermediate representation of source code as the code modality, i.e., OSCAR~\cite{peng2021oscar}. To avoid the influence of input modality on the comparison, we exclude OSCAR from the selected CodePTMs. For the natural language modality, these models cover all the three concrete types of inputs, namely, ``Doc'' (is the natural language description of the method), ``Post'' (refers to the natural language posts on Stack Overflow), and ``Names'' (indicate the sequence of function names present in the code). For the structure modality, these models also cover most frequently used ones, e.g., Data Flow Graph (DFG) and Abstract Syntax Trees (AST). To make the comparison sufficient, our selection covers all types of CodePTMs w.r.t. the number of input modalities they employ, i.e, unimodal, bimodal, and multimodal. 

\subsection{Pre-Training Tasks}
\label{section:codeptm_overview_objective}

In Table~\ref{table:codeptms}, we briefly present the pre-training tasks and their combination strategies of each model. Detailed definition of each pre-training task is given in Table~\ref{table:pretrainingtask}, where we also classify these tasks into four categories according to their input modalities: (1) Natural-Language-Aware or Code-Aware (\textbf{NLA/CA}) tasks, which are originated in NLP and can be applied to either NL or Code sequence to mine latent information from NL or Code; (2) Code-Aware Only (\textbf{CA Only}) tasks, which can only be applied to mine latent information from code text; (3) Structure-Aware Only (\textbf{SA Only}) tasks, which aim to learn representations of the code structure; and (4) Cross-Modal-Aware (\textbf{CMA}) tasks, which seek to acquire knowledge from multiple input modalities and are further subdivided into two categories based on which input modalities are involved, namely Code-NL (\textbf{CN}) and Code-Structure (\textbf{CS}).

Additionally, we mark the Objective (\textbf{O.}) category for each task and they are Generative (\textbf{G}, i.e., generate tokens) and Categorical (\textbf{C}, i.e., predict labels). In the empirical comparison, we would investigate the impacts of both types and objectives of pre-training tasks on the performance of pre-trained models on SE tasks.  

\subsection{Performance Discussion}
\label{section:codeptms_discussion}

As can be seen from Table 2, if only the SE tasks and corresponding datasets shown in Table 1 are considered, the TE-based models have no obvious tendency in applying to the Understanding or Generation tasks, while the TD- and TF-based ones are mostly applied to the Generation tasks. Since current  CodePTMs differ not only in architecture, input modality, and pre-training tasks, but also in their pre-training settings (such as datasets, model sizes, etc.), it is impossible to discuss the pure impacts of pre-training tasks on the performance of pre-trained models according to results in Table~\ref{table:codeptms}. Here, we present some findings derived from the comparison of performance of current CodePTMs on the SE tasks to which more than one CodePTMs are applied to. 

\subsubsection{Defect Detection} It seems that utilizing Cross-Modal-Aware pre-training tasks does not help to improve the performance on this task. Besides, it is not known that whether considering code structure to learn code representation would promote the performance on this task. 

\subsubsection{Clone Detection} Using only NL/Code Aware pre-training tasks can achieve the same SOTA performance as using complex pre-training tasks. But it seems that using structure-aware related pre-training tasks has a positive impact on the performance, since when using Structure-Aware Only and Code-Structure-Aware pre-training tasks, it is possible to obtain performance close to SOTA with the TE architecture. Shen et al.~\cite{shen-etal-2022-benchmarking} shows a similar option that simply taking a program as a token sequence is insufficient for modeling the program structure.

\subsubsection{Code Translation} Generally, additionally considering structure information can help learn better code representation, but on this task, using code structure as extra input plus adopting Code-Structure aware pre-training task (e.g., SPT-Code~\cite{niu2022sptcode} and GraphCodeBert~\cite{guo2021graphcodebert}) still cannot achieve SOTA performance. 

\subsubsection{Bug Fixing} Obviously, using NL as extra input is helpful to improve the performance of CodePTMs on the BF task. But it should be combined with appropriate pre-training tasks. While utilizing NL together with CA Only and CN aware pre-training tasks can achieve the SOTA performance. However, the use of structural information, including structure related pre-training tasks failed to bring a performance improvement. This finding is consistent with the that of Chirkova and Troshin~\cite{chirkova2021empirical}, where they show that the role of using code syntax structure depends on the specific task.

\subsubsection{Code Summarization} It is not surprising that the SOTA performance on this task is achieved by a model that uses both NL as extra input and a Code-NL aware pre-training task. But it is unexpected that the gap between the performance of the model that also use NL and Code-NL aware pre-training tasks and the SOTA is far greater than the gap between the performance of the models that do not use any NL-related information or pre-training tasks and the SOTA. Whether this is caused by the different experimental setups is worth exploring. Besides, it seems that pre-traininig tasks based on Method Names are useless compared to those utilizing Document. 

\subsubsection{Code Generation} Although the SOTA performance on the CG task is achieved by the model using NL as extra input and utilizing Code-NL aware pre-training task, it is not certain that the improvement is made by the NL input or the cross-modal pre-training task. The reason is that the SOTA model also adopts two Code Aware Only pre-training tasks, e.g., IT and SIMLM. This remains to be further explored through more experiments.

\section{Experimental Setup}
\label{section:experimental_setup}

\subsection{Re-pre-training Settings}
\label{section:setup_pre_training_dataset}

\subsubsection{Model Preparation} 
Since selected CodePTMs have different situations of open source code, we divide them into three cases for preparation:
\begin{itemize}
    \item For those having complete public source code (including pre-training and fine-tuning), we will set them to our experimental setups and then use the provided source code directly for our experiments. They are CugLM, MLM+DOBF, PLBART, T5-learning, CoTexT, TreeBERT, and SPT-Code.
    \item For those only giving the pre-trained model and the fine-tuning source code, we first re-implement the pre-training code or use the available checkpoint, and then fine-tune the pre-trained model. They are CuBERT, CodeBERT, GraphCodeBERT, ProphetNet-Code, and CodeT5.
    \item For those who do not have publicly available fully correct source code, we re-implement them based on their papers and available source code (if any). They are C-BERT, CodeGPT-adapted, and DeepDebug\footnote{To verify the correctness of our re-implemented models, we evaluate their performance according to the pre-training and fine tuning settings described in the original paper. Concrete results can be found in the supplementary file. Statistical significance tests show that the performance of our re-implemented models is not significantly different from the original reported ones.}.
\end{itemize}

\subsubsection{Unified Pre-training Dataset}

We choose CodeSearchNet~\cite{husain2020codesearchnet} as the pre-training dataset, which is the most used pre-trained dataset for CodePTMs (8 out of 15). The CodeSearchNet corpus is programmatically obtained by scraping open-source repositories and pairing individual functions with their (processed) documentation as natural language annotation. It includes more than 6.4M codes (of which more than 2.1M codes with paired documentation) from 6 PLs, i.e., Java, Python, JavaScript, PHP, Go and Ruby. Considering there are also some downstream tasks that also use CodeSearchNet, to prevent data leakage of the validation and test sets, we solely employ the training set of CodeSearchNet as the pre-training data. The data statistics are shown in Table~\ref{table:csn_statistics}. The ``filtered'' version~\cite{lu2021codexglue} is a subset obtained by filtering some low-quality documents on top of the code with documentation, which is used as the dataset for the downstream task of code summarization.

It is worth noting that PLBART uses ``Post'' as the natural language data, however, this is not available in CodeSearchNet. Therefore, we extract the ``Docs'' from it instead of ``Post'' as the natural language information in the pre-training data.

\begin{table}[t]
    \centering
    \caption{The CodeSearchNet corpus data statistics.}
    \label{table:csn_statistics}
    \resizebox{\linewidth}{!}{
    \begin{tabular}{lrrrrr}
        \toprule
            \multirow{2}{*}{\textbf{PL}} &
            \multirow{2}{*}{\textbf{All}} &
            \multirow{2}{*}{\textbf{w/ Doc}} &
            \multicolumn{3}{c}{\textbf{Filtered}} \\
        \cline{4-6}
            &
            &
            &
            \textbf{Train} &
            \textbf{Dev} &
            \textbf{Test} \\
        \midrule
            Java       & 1,569,889 & 500,754   & 164,923 & 5,183  & 10,955 \\
            Python     & 1,156,085 & 458,219   & 251,820 & 13,914 & 14,918 \\
            JavaScript & 1,857,835 & 143,252   & 58,025  & 3,885  & 3,291  \\
            PHP        & 977,821   & 662,907   & 241,241 & 12,982 & 14,014 \\
            Go         & 726,768   & 319,256   & 167,288 & 7,325  & 8,122  \\
            Ruby       & 164,048   & 52,905    & 24,927  & 1,400  & 1,261  \\
        \hline
            Total      & 6,452,446 & 2,137,293 & 908,224 & 44,689 & 52,561 \\
        \bottomrule
    \end{tabular}}
\end{table}

\subsubsection{Unified Model Size}
\label{section:setup_others}

The ``base'' version of Transformer~\cite{vaswani2017transformer} is used as the unified model scale. The dimensions of model and feed forward are set as 512 and 2048, respectively, the number of layers and self-attention heads are 6 and 8, and the dropout rate is 0.1. We also set the size of the vocabulary to 50,000$\pm$1\%\footnote{This is because the subword tokenizer constructs the word list by merging subwords, and it is not possible to set the word list size exactly at 50,000 before and after a certain merge, so we consider any variation of not more than one percent to be qualified, for example, CodeBERT's word list size is 50,265.}, and the input and output length limits are set to 512. 

\subsection{Fine-tuning Settings}

For TE-, TD- and TF-based models, different settings are needed to apply them to code understanding and code generation tasks respectively. When TE-based models are applied to the understanding task, we extract the vector in the output states that correspond to the special symbols (e.g., [CLS] and \textless s\textgreater) that represent the classification position in the input. In contrast, when they are applied to the generation task, we randomly initialize a decoder of the same size to form an encoder-decoder architecture with the model. In the case of TD-based models, we use the last hidden state to indicate the prediction when they are applied to understanding tasks, and apply them to generation under the auto-regressive fashion without any modification.

To apply the TF-based model to the code understanding task, we use the settings of their base model, which is also consistent with the settings in their papers. Namely, for the BART-based models (PLBART and SPT-Code), we use the last step of the hidden states of the decoder, and for the T5-based models (i.e., DeepDebug, T5-learning, CoTexT and CodeT5), we turn the task into a text-to-text form by converting each label to text. In addition, if a representation vector for the T5-based model is needed, we use the same settings as for BART-based models since the text-to-text setting cannot be applied to these tasks. For other models, since they do not specify the settings to be applied to the code understanding task, we use the BART settings by default. Whereas with respect to code generation tasks, the TF-based model can be applied directly without any modification.

For each SE task, we fine-tune the pre-trained models with its training set and perform a validation on its development set at the end of each epoch. The limit of epochs for fine-tuning is set to 50\footnote{We do not set the number of fine-tuning epochs too large because the pre-trained model is supposed to be adapted to the given downstream task in a shorter time. Moreover, setting the number of fine-tuning epochs too large may cause the models to forget the knowledge learned during pre-training and degrade from generic to task-specific knowledge. Since these models are almost identical in terms of architecture and under the same experimental setup, this would lead to very similar performance between the models, and it would not be possible to compare the pre-training techniques proposed by each model based on the results.}. The validation metric is the first evaluation metric shown in Table~\ref{table:se_tasks}. We execute an early stop when the validation metric does not set a new best record for 20 consecutive validations. For fine-tuning each pre-trained model for each SE task, we experiment three times using different random seeds (i.e., 24, 42 and 81) and report the mean of testing results. As to the optimizer for fine-tuning, we use AdamW for all tasks. For each time fine-tuning, we tune hyper-parameters using grid search, where we select learning rate from \{1e-5, 3e-5, 5e-5, 1e-4\}, warm-up steps from \{500, 1000, 2000\}, and batch size per device from \{16, 32, 64\}.

\subsection{Other Settings}

All data pre-processing procedures and input representations during pre-training and fine-tuning are performed strictly according to the source code (if available) or the methods described in the paper. We perform a total of 500K steps of pre-training. For other hyperparameter settings during pre-training, namely optimizer, initial learning rate, batch size, warmup steps, we set them to the same settings as in the original text or the default settings in the publicly available source code. If no relevant settings are found in either, we will default to the following settings.

As the default settings, we use AdamW~\cite{loshchilov2018decoupled} as our optimizer, the initial learning rate is 5e-5 and the warmup step is set as 2000. All experiments, including pre-training and fine-tuning, are performed on 4$\times$NIVDIA Tesla V100s. For pre-training, the default batch size is 256. 

\section{Re-pre-training Speed of CodePTMs}
\label{section:experimental_results}

Since we re-run the pre-training process for all CodePTMs except CodeT5, it make us feasible to compare their pre-training speed. We calculate the average seconds they took per 1K pre-training steps and present the results in Table~\ref{table:pre_training_speed}. Incorporating Table~\ref{table:codeptms} and Table~\ref{table:pre_training_speed}, we find that several points deserve mention. 

First, CodePTMs of highest pre-training speed are TE-based and TD-based models, while the lowest ones are TF-based models. That makes sense since that the TF-based model needs to first pass the encoder input to its encoder, and then pass the encoder output and the decoder input to its decoder. If it is a 6-layer Transformer, then actually the TF-based model needs to make a 12-layer pass, and because their decoder has three sub-layers, more than the two sub-layers of encoder, the pre-training time of TF-based models is theoretically more than twice as long as encoder or decoder-only.

The length of the input is also an important factor. Although we unify all input lengths and use padding symbols for padding where they are insufficient, different CodePTMs do not have the same input length due to the different modalities in their inputs. For example, with both the same architecture and pre-training tasks, CoTexT with inputs consisting of code and documentations concatenated together has a lower training speed than DeepDebug with inputs containing only code, i.e., DeepDebug has a shorter input sequence than CoTexT. And the slowest model, SPT-Code, is the only TF-based model which uses three modalities.

Besides, the pre-training objective also influences the pre-training speed since it affects the length of the output during pre-training. Specifically, it costs more time for objectives prediting longer output sequence. This mainly related to the TD- and TF-based models. For example, PLBART, T5-learning and ProphetNet-Code are all TF-based models and have only single pre-traning objective, but it is found that PLBART is much slower compared to T5-learning and ProphetNet-Code. The reason is that DAE needs to make the model regenerate the complete input sequence, while SMLM and SNP only need to generate several sub-sequences in the input. 

Finally, additional processing/encoding of input before encoding also lower down the pre-training speed. Taking the comparison of TreeBERT and PLBART as an example, which are both TF-based and generating the entire code sequence, TreeBERT is slower than PLBART although the AST Paths used in TreeBERT is shorter than the Post used in PLBART. The reason is that in order to get the AST Paths, TreeBERT needs to embed the nodes on each path and then perform a series of concatenations and other operations.

\begin{table}[t]
    \centering
    \caption{Pre-Training Speed of re-run CodePTMs.}
    \label{table:pre_training_speed}
    \resizebox{\linewidth}{!}{%
        \begin{tabular}{lr|lr}
            \toprule
                \textbf{CodePTM} & \textbf{Sec. / 1K Steps} & \textbf{CodePTM} & \textbf{Sec. / 1K Steps} \\
            \midrule
                CuBERT          & 1,330.31 & T5-learning     & 3,097.34 \\
                CodeBERT        & 1,459.37 & PLBART          & 3,101.92 \\
                C-BERT          & 1,388.23 & ProphetNet-Code & 3,092.28 \\
                GraphCodeBERT   & 1,502.06 & DeepDebug       & 3,136.73 \\
                CugLM           & 1,351.52 & CoTexT          & 3,192.51 \\
                MLM+DOBF        & 3,053.36 & TreeBERT        & 3,167.83 \\
                CodeGPT-adapted & 1,314.64 & SPT-Code        & 3,216.49 \\
            \bottomrule
        \end{tabular}%
    }
\end{table}

\section{Performance of Re-pre-trained CodePTMs}

We present the results of all the re-pre-trained CodePTMs for the code understanding and generation tasks in Tables~\ref{table:results_und} and \ref{table:results_gen}, respectively. In the two result tables, we divided these models into three groups according to their architectures, namely, TE-based, TD-based, and TF-based. All models are arranged in the same order as in Table~\ref{table:codeptms}. The best scores of the TE-based group and TF-based groups are all in bold. The second best scores of the TF-based group are underlined.

\begin{table}[t]
    \centering
    \caption{Experimental results on code understanding tasks.}
    \label{table:results_und}
    \resizebox{\linewidth}{!}{%
        \begin{tabular}{lccccccc}
            \toprule
                \multirow{2}{*}{Model} & 
                DD & 
                \multicolumn{3}{c}{CD} & 
                CR & 
                CS & 
                QA \\ 
            \cline{3-5}
                & 
                Acc & 
                F1 & 
                P & 
                R & 
                MAP & 
                MRR & 
                Acc \\
            \midrule
            C-BERT &       
                57.54 & 
                95.72 & 
                \textbf{97.70} & 
                93.83 &       
                79.88 &       
                26.17 &       
                43.46 \\
            CuBERT &
                57.07 & 
                94.86 & 
                95.88 & 
                93.86 &       
                79.45 &       
                25.64 &       
                42.46 \\
            CugLM &       
                \textbf{58.15} & 
                \textbf{96.21} & 
                95.73 & 
                \textbf{96.69}  &       
                80.67 &       
                27.63 &      
                48.66 \\
            CodeBERT &       
                57.59 & 
                95.50 & 
                96.21 & 
                94.81 &       
                80.78 &       
                26.96 &       
                49.24 \\
            GraphCodeBERT &       
                \underline{57.82} & 
                \underline{96.12} & 
                \underline{96.89} & 
                \underline{95.35} &
                \textbf{83.51} &      
                \textbf{33.38} &      
                \textbf{52.14} \\ 
        \hline \hline
            CodeGPT-adapted &       
                59.19 & 
                96.10 & 
                97.15 & 
                95.08 &       
                79.16 &       
                31.71 &       
                50.69 \\ 
        \hline \hline
            DeepDebug &       
                58.48 & 
                96.02 & 
                94.69 & 
                \underline{97.39} &       
                80.67 &       
                30.72 &       
                50.87 \\
            MLM+DOBF &       
                58.89 & 
                \textbf{97.12} & 
                96.93 & 
                97.31 &       
                80.98 &       
                28.47 &       
                50.79 \\
            PLBART &       
                \underline{59.23} & 
                96.24 & 
                94.74 & 
                \textbf{97.79} &       
                82.99 &       
                32.23 &       
                51.65 \\
            T5-learning &       
                58.98 & 
                96.67 & 
                \underline{97.20} & 
                96.15 &       
                82.65 &       
                30.35 &       
                50.73 \\
            ProphetNet-Code &       
                58.35 & 
                96.18 & 
                94.65 & 
                \textbf{97.76} &       
                81.51 &       
                31.91 &       
                51.04 \\
            CoTexT &       
                58.93 & 
                96.45 & 
                96.42 & 
                96.47 &       
                82.44 &       
                32.68 &       
                \underline{51.81} \\
            CodeT5 &       
                58.92 & 
                \underline{96.73} & 
                96.50 & 
                96.97 &       
                \underline{83.11} &       
                \textbf{33.10} &      
                51.28 \\
            TreeBERT &       
                \textbf{59.69} & 
                \textbf{97.05} & 
                \textbf{97.56} & 
                96.55 &      
                \textbf{83.48} &       
                32.36 &      
                \textbf{52.09} \\
            SPT-Code & 
                58.85 & 
                96.15 & 
                95.89 & 
                96.41 &
                82.78 &       
                \underline{32.96} &       
                50.63 \\
            \bottomrule
        \end{tabular}%
    }
\end{table}
\begin{table*}[htbp]
    \centering
    \caption{Experimental results on code generation tasks.}
    \label{table:results_gen}
    \resizebox{\textwidth}{!}{%
        \begin{tabular}{lcccccc|cccccc|c|cccccc|ccc}
            \toprule
                \multirow{3}{*}{\textbf{Model}} &
                \multicolumn{6}{c|}{\textbf{Code Translation}} &
                \multicolumn{6}{c|}{\textbf{Bug Fixing}} &
                \multirow{2}{*}{\textbf{CC}} &
                \multicolumn{6}{c|}{\textbf{Code Summarization}} &
                \multicolumn{3}{c}{\multirow{2}{*}{\textbf{\begin{tabular}[c]{@{}c@{}}Code\\ Generation\end{tabular}}}} \\
            \cline{2-13}\cline{15-20}
                &
                \multicolumn{3}{c}{\textbf{Java-\textgreater{}C\#}} &
                \multicolumn{3}{c|}{\textbf{C\#-\textgreater{}Java}} &
                \multicolumn{3}{c}{\textbf{small}} &
                \multicolumn{3}{c|}{\textbf{medium}} &
                &
                \textbf{Java} &
                \textbf{Python} &
                \textbf{JS} &
                \textbf{PHP} &
                \textbf{Go} &
                \textbf{Ruby} &
                \\
            \cline{2-23}
                &
                \textbf{EM} &
                \textbf{B.} &
                \textbf{C.B.} &
                \textbf{EM} &
                \textbf{B.} &
                \textbf{C.B.} &
                \textbf{EM} &
                \textbf{B.} &
                \textbf{C.B.} &
                \textbf{EM} &
                \textbf{B.} &
                \textbf{C.B.} &
                \textbf{EM} &
                \textbf{B.} &
                \textbf{B.} &
                \textbf{B.} &
                \textbf{B.} &
                \textbf{B.} &
                \textbf{B.} &
                \textbf{EM} &
                \textbf{B.} &
                \textbf{C.B.} \\ 
            \midrule
            C-BERT &
                \underline{58.70} &
                78.06 &
                83.11 &
                \textbf{59.56} &
                \textbf{72.43} &
                79.34 &
                15.44 &
                72.41 &
                \underline{73.63} &
                4.98 &
                86.81 &
                \textbf{84.69} &
                72.30 &
                \underline{16.41} &
                \underline{17.97} &
                13.01 &
                22.02 &
                16.93 &
                10.76 &
                17.44 &
                33.65 &
                37.13 \\
            CuBERT &
                57.35 &
                78.25 &
                83.53 &
                56.24 &
                71.00 &
                78.13 &
                15.89 &
                73.01 &
                72.33 &
                5.17 &
                86.48 &
                \underline{84.15} &
                72.26 &
                15.92 &
                17.68 &
                12.95 &
                21.97 &
                16.61 &
                10.64 &
                17.20 &
                32.52 &
                36.79 \\
            CugLM &
                \textbf{59.53} &
                \textbf{80.40} &
                \underline{84.81} &
                \underline{58.95} &
                70.61 &
                78.45 &
                15.06 &
                \underline{73.69} &
                \textbf{74.46} &
                \underline{6.50} &
                86.36 &
                83.15 &
                \textbf{74.23} &
                16.01 &
                17.67 &
                13.82 &
                22.26 &
                \underline{17.24} &
                11.05 &
                \textbf{18.50} &
                \textbf{35.88} &
                \textbf{38.04} \\
            CodeBERT &
                57.17 &
                79.72 &
                83.79 &
                55.49 &
                70.82 &
                \underline{79.40} &
                \underline{16.42} &
                72.81 &
                72.75 &
                5.61 &
                \textbf{87.60} &
                83.30 &
                72.12 &
                \textbf{16.78} &
                \textbf{18.13} &
                \underline{13.93} &
                \textbf{23.45} &
                17.15 &
                \underline{11.28} &
                17.76 &
                33.95 &
                37.22 \\
            GraphCodeBERT &
                58.56 &
                \underline{80.34} &
                \textbf{84.85} &
                58.87 &
                \underline{71.55} &
                \textbf{79.70} &
                \textbf{16.70} &
                \textbf{75.56} &
                72.80 &
                \textbf{7.92} &
                \underline{87.13} &
                83.80 &
                \underline{72.64} &
                \textbf{16.76} &
                \textbf{18.11} &
                \textbf{13.96} &
                \underline{22.65} &
                \textbf{17.43} &
                \textbf{11.52} &
                \underline{18.02} &
                \underline{35.45} &
                \underline{37.56} \\
        \hline \hline
            CodeGPT-adapted &
                60.61 &
                81.96 &
                85.34 &
                62.12 &
                75.38 &
                83.29 &
                17.82 &
                75.32 &
                76.70 &
                7.50 &
                86.09 &
                84.54 &
                75.42 &
                16.81 &
                18.46 &
                14.01 &
                22.75 &
                17.48 &
                10.78 &
                19.22 &
                32.86 &
                35.03 \\
        \hline \hline
            DeepDebug &
                \underline{63.67} &
                82.31 &
                \underline{86.63} &
                \underline{64.06} &
                73.63 &
                81.67 &
                18.06 &
                75.53 &
                74.40 &
                \underline{9.80} &
                89.88 &
                84.89 &
                73.13 &
                17.89 &
                18.06 &
                13.41 &
                23.02 &
                17.27 &
                12.99 &
                18.78 &
                35.87 &
                39.18 \\
            MLM+DOBF &
                60.07 &
                81.94 &
                85.90 &
                61.58 &
                73.66 &
                80.98 &
                16.10 &
                75.39 &
                74.65 &
                7.96 &
                86.62 &
                84.30 &
                \underline{73.52} &
                17.00 &
                18.31 &
                14.26 &
                22.87 &
                17.52 &
                11.46 &
                17.53 &
                35.22 &
                37.81 \\
            PLBART &
                \textbf{64.64} &
                \underline{83.76} &
                86.41 &
                \textbf{65.43} &
                \textbf{76.18} &
                \textbf{83.92} &
                17.35 &
                76.04 &
                76.32 &
                9.06 &
                \textbf{91.63} &
                \textbf{86.91} &
                73.10 &
                17.52 &
                18.36 &
                14.68 &
                23.66 &
                17.98 &
                13.13 &
                17.65 &
                35.76 &
                37.46 \\
            T5-learning &
                60.46 &
                82.78 &
                84.00 &
                61.56 &
                74.69 &
                82.61 &
                16.58 &
                74.28 &
                75.60 &
                7.30 &
                87.58 &
                85.62 &
                73.45 &
                17.30 &
                \underline{18.77} &
                14.37 &
                23.04 &
                17.72 &
                12.45 &
                18.39 &
                36.24 &
                38.83 \\
            ProphetNet-Code &
                62.45 &
                81.97 &
                85.01 &
                63.61 &
                \underline{75.94} &
                82.11 &
                17.92 &
                75.59 &
                74.01 &
                8.75 &
                90.72 &
                86.12 &
                \textbf{74.61} &
                \underline{18.50} &
                17.93 &
                14.69 &
                23.63 &
                17.45 &
                13.44 &
                18.92 &
                \underline{36.63} &
                39.57 \\
            CoTexT &
                61.23 &
                83.10 &
                84.12 &
                62.09 &
                71.69 &
                80.88 &
                \underline{20.15} &
                77.16 &
                76.46 &
                9.33 &
                \underline{91.51} &
                85.68 &
                72.87 &
                18.20 &
                18.59 &
                14.63 &
                23.88 &
                18.19 &
                13.22 &
                \underline{19.11} &
                36.57 &
                \underline{40.59} \\
            CodeT5 &
                62.93 &
                82.32 &
                \textbf{86.68} &
                63.38 &
                74.20 &
                \underline{83.65} &
                \textbf{20.36} &
                \underline{77.43} &
                \textbf{76.75} &
                \textbf{9.83} &
                91.05 &
                \underline{86.70} &
                73.39 &
                \textbf{18.57} &
                \textbf{18.98} &
                \textbf{15.18} &
                \textbf{24.19} &
                \textbf{18.55} &
                \textbf{13.88} &
                \textbf{19.92} &
                \textbf{37.09} &
                \textbf{41.16} \\
            TreeBERT &
                62.74 &
                82.64 &
                85.76 &
                62.16 &
                73.06 &
                81.85 &
                20.00 &
                75.79 &
                75.13 &
                9.28 &
                89.81 &
                85.17 &
                72.50 &
                17.65 &
                17.72 &
                13.87 &
                23.31 &
                17.94 &
                12.78 &
                18.74 &
                36.38 &
                39.68 \\
            SPT-Code &
                62.76 &
                \textbf{83.98} &
                85.93 &
                62.78 &
                75.41 &
                82.44 &
                19.52 &
                \textbf{77.90} &
                \underline{76.54} &
                9.07 &
                90.41 &
                86.36 &
                73.08 &
                18.28 &
                18.31 &
                \underline{14.85} &
                \underline{23.92} &
                \underline{18.29} &
                \underline{13.58} &
                19.05 &
                36.59 &
                40.49 \\
            \bottomrule
        \end{tabular}%
    }
\end{table*}

First of all, since T5-learning in the original paper uses the same model size as our setup, we first compare the scores of T5-learning reported in the original paper and our results. In the original paper, the pre-training dataset of T5-learning only uses the Java subset of the CodeSearchNet~\cite{mastropaolo2021t5-learning}, and the pre-training dataset of our paper uses the whole CodeSearchNet. By comparing the results of T5-learning on the Bug Fixing downstream task (i.e., the EMs on small/medium are 10/3 in the original paper and 16.58/7.30 in our results), we find that despite the fact that the programming language of the extra pre-training data is different from the downstream tasks, these extra data can still improve the model's performance on this downstream task. This finding is consistent with Niu et al.~\cite{niu2022sptcode}.

Next, we will revisit the discussion, which differs from the discussion in subsection \textbf{Performance Discussion} in two ways: (1) this discussion is more comprehensive, as we experimentally obtained results for the full set of CodePTMs on all types of tasks, and (2) this discussion will focus on the impacts of pre-training tasks on the performance of pre-trained models on SE tasks on the premise of the same model size, since we have ensured that the experimental settings such as pre-training datasets and evaluation datasets are unified. We will present a complete and constructive discussion, by analyzing each task, in turn. Please note that for each SE task, the performance of all pre-trained models on it are verified significantly different using Approximate Randomization tests\footnote{\url{https://github.com/danieldk/approx-rand-test}} where the \textit{P-value}s are less than 0.05. 

Before we go further into the discussion of the experimental results, we present at-a-glance view of the results in Table~\ref{table:best_pre_training_task}, i.e., the relations among the investigated pre-training tasks and downstream SE tasks. we first count the top five performing models for each downstream task. Then we determine the pre-training tasks utilized by these models and present the most frequently occurring combinations in Table~\ref{table:best_pre_training_task}, where CA, SA, CN, CS means code-aware only, structure-aware only, code-NL-aware, and code-structure-aware pre-training tasks, respectively. From the Table~\ref{table:best_pre_training_task}, we have some takeaways.
\begin{enumerate}
    \item Structure-aware pre-training tasks can be beneficial on most of downstream tasks, except Code Translation, Code Complition and Code Generation.
    \item MLM and its variants, such as SMLM, are a good choice for most downstream tasks but Code Completion.
    \item Code-NL-aware tasks become powerful when it comes to generation downstream tasks, excluding Code Completion.
    \item FLM is best suitable for Code Completion.
\end{enumerate}

\begin{table}[t]
    \centering
    \footnotesize
    \caption{The best performing pre-training tasks on each task.}
    \label{table:best_pre_training_task}
    \begin{tabular}{ll|ll}
        \toprule
            Task & Pre-training tasks & Task & Pre-training tasks \\
        \midrule
            DD   & SA, SMLM, DAE      & CT   & DAE, SMLM, CN      \\
            CD   & MLM, SA, SMLM      & BF   & SMLM, SA, CN       \\
            CR   & SA, MLM, CS        & CC   & FLM                \\
            CS   & SA, MLM, CN        & SM   & SMLM, CN, CA, SA   \\
            QA   & SA, MLM, SMLM      & CG   & SMLM, CA, CN       \\
        \bottomrule
    \end{tabular}
\end{table}

\subsection{SE Task Analysis}
\label{section:discussion_task}

Unrelated to the comparison of performance between CodePTMs, we begin with a brief analysis and discussion of all tasks by looking at the performance levels of all CodePTMs on them. First, we can see that for the code understanding task, all models reach a high level on the clone detection task, suggesting that the BigCloneBench dataset does not pose much difficulty for CodePTMs and that a more difficult dataset is needed to distinguish between them. And Code-to-Code Retrieval needs to consider the same issues. In contrast, Code Search presents a greater challenge for CodePTMs, suggesting that the filtered AdvTest does present an increased difficulty compared to using CodeSearchNet directly (where CodePTMs report results greater than 70). For the two remaining tasks, Defect detection and Code Question Answering, they remain moderately difficult in comparison.

As for the code generation task, Code Translation, Completion are of moderate difficulties. Bug Fixing, on the contrary, has a particularly low EM on both subsets, and the way the subsets are divided (i.e., the codes in small are all below 50 in length and the codes in medium are all between 50 and 100 in length) results in all CodePTMs having a smaller EM on medium than they do on small. And the reason for the high scores of BLEU and CodeBLEU is that Bug Fixing only repairs the bug in the code, while the rest is copied as is. At this point, the longer length of the code allows more n-gram matching opportunities for the sequences predicted by the model, resulting in a higher BLEU and CodeBLEU on the medium subset than on small.

The Code Summarization and Generation are also challenging, with the CodePTMs scoring at different levels on different languages in the code summarization. Through our analysis, and by referring to the results on this dataset in other papers, we argue that this is due to the different difficulty of the dataset on each language.

\subsection{Code Understanding Tasks}
\label{section:discussion_understanding}

\subsubsection{Defect Detection}

FLM has better results on this task. Specifically, for TE-based models, we find that combining FLM and MLM (CugLM) possesses better performance than using only MLM (C-BERT) or other pre-training tasks combined with MLM (CuBERT, CodeBERT, GraphCodeBERT). The effectiveness of FLM is reinforced by the performance of CodeGPT-adapted, where using only FLM allowed it to achieve a higher score than all TE-based models. Furthermore, it seems that using only code, without natural language and structural information can achieve good results on this task, when the model is based on TE or TD.

However, the best-performing TF-based model, TreeBERT, which is the best-performing of all models as well, also learns code-related features, i.e., structure information, for its pre-training tasks, i.e., TMLM and NOP. Besides, TreeBERT's best performing also indicates that (1) it is better to learn code-specific characteristics rather than just treating code as natural language, and (2) TF can achieve a level similar to TE and TD on the understanding task.

\subsubsection{Clone Detection}

In contrast to defect detection, MLM and its variants show strong positive impact on this task. The best TE-based and TF-based performers in the three metrics are CugLM, C-BERT, and MLM+DOBF, TreeBERT, PLBART. CugLM and C-BERT both adopt MLM as their pre-training tasks. For three TF-based models, they use MLM, SMLM, TMLM, which are all based on MLM.

\subsubsection{Code-to-Code Retrieval}

This task requires the model to identify code snippets that are different in form but functionally identical. Therefore, simply borrowing pre-training tasks from NLP does not help the model to achieve good results, such as C-BERT, CodeGPT-adapted, DeepDebug, etc. Instead, by learning code-specific features, the model can have a better ability to discriminate between different forms but the same functional code. GraphCodeBERT, for example, achieves the best results among all models by using EP and NA to learn structure information from DFGs. In addition, the best TF-based model, TreeBERT, learns structure information from ASTs by utilizing TMLM and TEP.

\subsubsection{Code Search and Question Answering}

We discuss these two tasks together because, in addition to both being understanding tasks, they are both tasks that span both code and natural language modalities, and both require models to have alignment capabilities between code and documentation. Thus, we can see that models that learn both code and natural language during pre-training, such as CodeBERT, GraphCodeBERT, PLBART, CoTexT, CodeT5, SPT-Code, etc., all achieve relatively higher scores.

Furthermore, when comparing GraphCodeBERT and CodeBERT, we can see that the performance of the former is significantly improved compared to the latter due to the inclusion of code-specific learning in the pre-training task. This indicates that code-specific features such as structure information, for example, also play a non-negligible role in these two tasks. This is also evidenced by the performance of TreeBERT and SPT-Code.

\subsection{Code Generation Tasks}

\subsubsection{Code Translation (CT)}

By employing either the TE or TF architecture, even when exclusively utilizing generative pre-training tasks such as C-BERT's MLM, CugLM's FLM, or PLBART's DAE, it is possible to achieve state-of-the-art (SOTA) performance on certain metrics. However, under the TF architecture, different from the TE architecture, more complex generative tasks are required to achieve SOTA performance. For example, since SMLM (which is randomly masking independent tokens and generating them) is simpler than DAE or FNM (which are dedicated to generate a consecutive sequence of tokens), T5-learning that using SMLM only fails to surpass Prophet-net or PLBART to achieve SOTA on any metric when other settings (e.g., architecture, input modality, and model size, etc.) are completely the same. 

Another interesting observation is that under TE architecture, using categorical tasks may has a negative impact. For example, integrating C-BERT with the NSP task (i.e. CuBERT) produces a significant drop on 4 out of 6 results. However, if the categorical task utilizes the structural information, a certain positive impact can be seen. For example, on the basis of C-BERT, after adding the EP and NA tasks (i.e., GraphCodeBERT), it achieves significant improvements on three metrics and two of them reach SOTA. Under TF architecture, it can be determined that using structural information-related tasks is helpful to improve the performance, such as SPT-Code. However, they must be utilized together with NLA/CA or CA Only tasks to be more effective. For example, TreeBERT using only SA Only and CS Aware tasks did not achieve SOTA results on any metrics and the reason is that it discards all tasks purely for code sequences.

Finally, comparing TE-based models and TD-based model (whose sizes are the same during pre-training), it proofs that on the code translation task, the generative pre-training tasks will be more effective if used in conjunction with architectures that are better suited for generative tasks, e.g., TD and TF. 

\subsubsection{Bug Fixing (BF)}

Although bug fixing is a task of generating complete code jusk like code translation task, its relation with pre-training tasks is a little different from that of code translation. First, while BF also relies on generative pre-training tasks, it benefits more from fine-grained tasks (i.e., dedicated to generate non-consecutive tokens, such as MLM, FLM, and DAE) than coarse-grained ones (i.e., dedicated to generate consecutive tokens sequence, such as FNP). For example, with TF architecture, ProphetNet-Code using FNP achieves not any SOTA on BF while PLBART using DAE makes two SOTA results. Second, categorical pre-training tasks positively contribute to improve the performance of pre-trained models on BF, especially token-level or node-level categorical tasks (e.g., RTD, EP, NA, and IT). For example, GraphCodeBERT using EP and NA achieves SOTA on three metrics. In contrast, the improvement of CuBERT (which uses the sentence-level task NSP) over C-BERT is not obvious. Last, for TE architecture, it seems that the structure aware pre-training tasks is a bit more powerful for BF than for CT, since the model using the structural aware tasks achieves more SOTA on BE than on CT. 

These findings make sense considering the characteristics of BF which essentially has two steps, first to locate bugs, and then to fix bugs. Locating bugs is more of a categorical task, while remediation is a generative task. Besides, because the bugs are generally at the token level, for both generative and categorical pre-training tasks, the fine-grained ones should have a greater contribution than the others on BF.

\subsubsection{Code Completion (CC)}

After complementing the performance of all TE- and TF-based models on the CC task, it is not surprising to find that the SOTA is obtained by the models using the FLM and FNP pre-training tasks, respectively. The reason is that CC task is to predict subsequent code tokens based on the given code as a prompt, which is similar to FLM or FNP. But unexpectedly, theoretically FNP is closer to CC task than FLM, but ProphetNet-code (TF-based) using FNP is surpassed by CodeGPT-adapted (TD-based) using FLM  even that its model size is twice of that of CodeGPT-adapted's. Given this, using natural language as additional pre-training input is the only explanation for ProphetNet-code's lag. This proves that, for CC tasks, integrating natural language has a negative impact on the model performance. However, whether using FNP to pre-train code under the TF architecture would achieve the new SOTA for CC task remains as an open question which requires further experimental verification.

\subsubsection{Code Summarization (SM)}

Compared with Table~\ref{table:codeptms}, there are three important findings derived from Table~\ref{table:results_gen} worth discussing. First of all, for both TE- and TF-based models, there are still cases where applying NLA/CA pre-training tasks to the code only surpasses applying them to both the code and the natural language. For example, CugLM outperforms CodeBERT on the datasets of both JS and Go languages. This proofs that for Code Summarization, the usage of natural language is important, but utilizing natural language with what kind of pre-training tasks is even more important. Another interesting clue is that CodeGPT-adapted (TD-based) pre-trained on code only outperforms TE-based SOTA models on 5 out of the 6 datasets. Second, under the same experimental settings, the MASS pre-training task based on Method Name can outperform the other generative pre-training tasks based on Document (e.g., SPT-Code achieves the second best scores on 5 language). Finally, theoretically speaking, structural information is helpful in learning better code semantic representation, but practical results in Table~\ref{table:results_gen} show that pre-training tasks designed for structural information have no significant contribution on code summarization tasks. 

\subsubsection{Code Generation (CG)}

One of the most remarkable findings is that CG aims to generate code from natural language input. However, it is fascinating to note that under the TE architecture, the model that attains state-of-the-art (SOTA) performance, namely CugLM, does not rely on any pre-training tasks associated with natural language (NL) nor does it utilize NL as input. Under the TF architecture, CodeT5 achieves the SOTA effect unsurprisingly, because to generate code based on NL is just one subtask of its BDG task. But it is worth mentioning that the second best results are achieved by CoTexT which uses simple generative pre-training task on both code and NL. This shows that NL is useful for CG tasks. Considering this, for TE architecture, it is likely that the pre-training task FLM used by CugLM is more suitable for the CG task than MLM. Moreover, ProphetNet-Code under TF architecture using FNP which is closer to CG than FLM fails to achieve better results than CoTexT, probably because it does not input code and NL together in a same data instance. In conclusion, that whether using FNP to pre-training both code and NL in a same data instance under TF architecture would achieve better performance than CoTexT, needs to be further explored through experiments.

\section{Threats to Validity}
\label{section:threats}

\subsubsection*{Construct Validity}

We know from Section~\ref{section:setup_pre_training_dataset} that our code usage varies for different CodePTMs. Except for the third case, i.e. CodePTMs where all the code is re-implemented by us, we cannot guarantee that the code provided by other CodePTMs will bring the same efficiency and effectiveness as our code. However, we keep our re-implemented model with the same settings as the original model for experiments and compare the results of our model and results in the original paper. Then we find that there were no significant differences between our re-implemented model and the original model under the t-test ($p<005$).

\subsubsection*{Internal Validity}

It is widely agreed that hyperparameters have a significant impact on the performance of deep learning models. But in pre-training phase, we are, as we declared in Section~\ref{section:setup_others}, using the hyperparameter settings provided in the original paper or source code; if not, we will use the default values. This poses some problems, firstly we are not sure if the original hyperparameter settings will give the best results for the current CodePTM, and secondly the default settings may have different effects for different CodePTMs. Finally, we standardize the scale of all models in this paper to the ``base'' version of Transformer, which may result in our results not being as good as the original model in the original text.

In addition, we have pre-trained all models with a single training budget of 500K steps. However, the models can surpass each other at different training budgets. Therefore, the single pre-training budget.

However, our aim in this paper is to exclude the bias caused by model scale. We therefore believe that despite although our setup will not give the best results, it does not introduce a threats to the findings and conclusions in this paper that is worth considering.

\subsubsection*{External Validity}

We can only guarantee that the conclusions in this article apply only to the tasks and datasets we have selected. Since different tasks require different capabilities of CodePTMs, and different datasets will have different data distributions, we cannot be quite sure if these conclusions will be applied to other tasks and datasets.

\section{Conclusion}
\label{section:conclusion}

This paper presents a comprehensive and fair experimental comparison of pre-training tasks used in current CodePTMs by re-pre-training and evaluating CodePTMs under a unified setting. By giving an overview of current CodePTMs, experimental results, and detailed analysis and discussion on relative strengths and weaknesses of pre-training tasks, we hope to provide some constructive inspiration for subsequent researchers, and also hope that more powerful CodePTMs will emerge in the near future.

\section*{Acknowledgments}
This work was supported by National Natural Science Foundation of China (61802167).
 
%

\bibliographystyle{IEEEtran}
\bibliography{refs}

%

\begin{IEEEbiography}[{\includegraphics[width=0.9in,height=1.1in,clip,keepaspectratio]{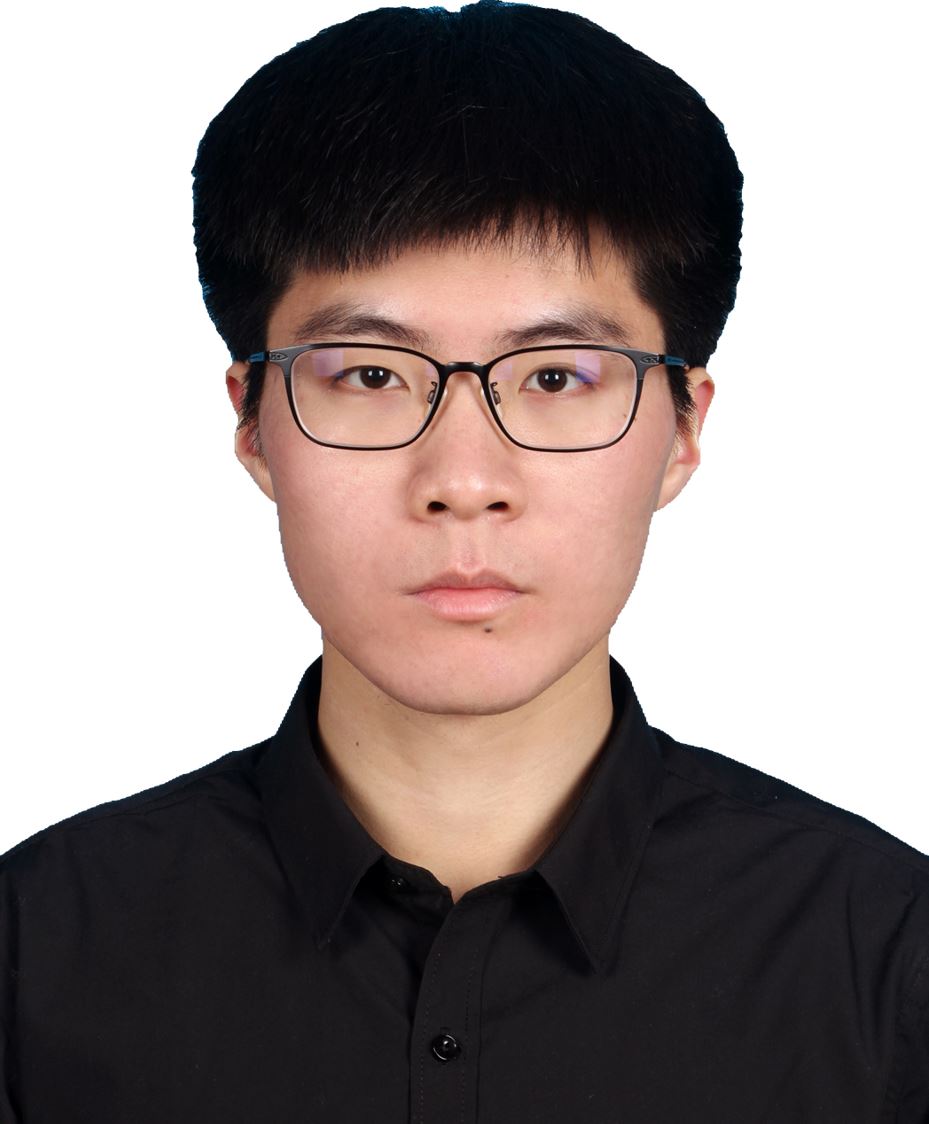}}]{Changan Niu}
is a Ph.D. candidate at Software Institute, Nanjing University, China. His research interests include software engineering, language models of source code, etc. 
\end{IEEEbiography}

\begin{IEEEbiography}[{\includegraphics[width=0.9in,height=1.1in,clip,keepaspectratio]{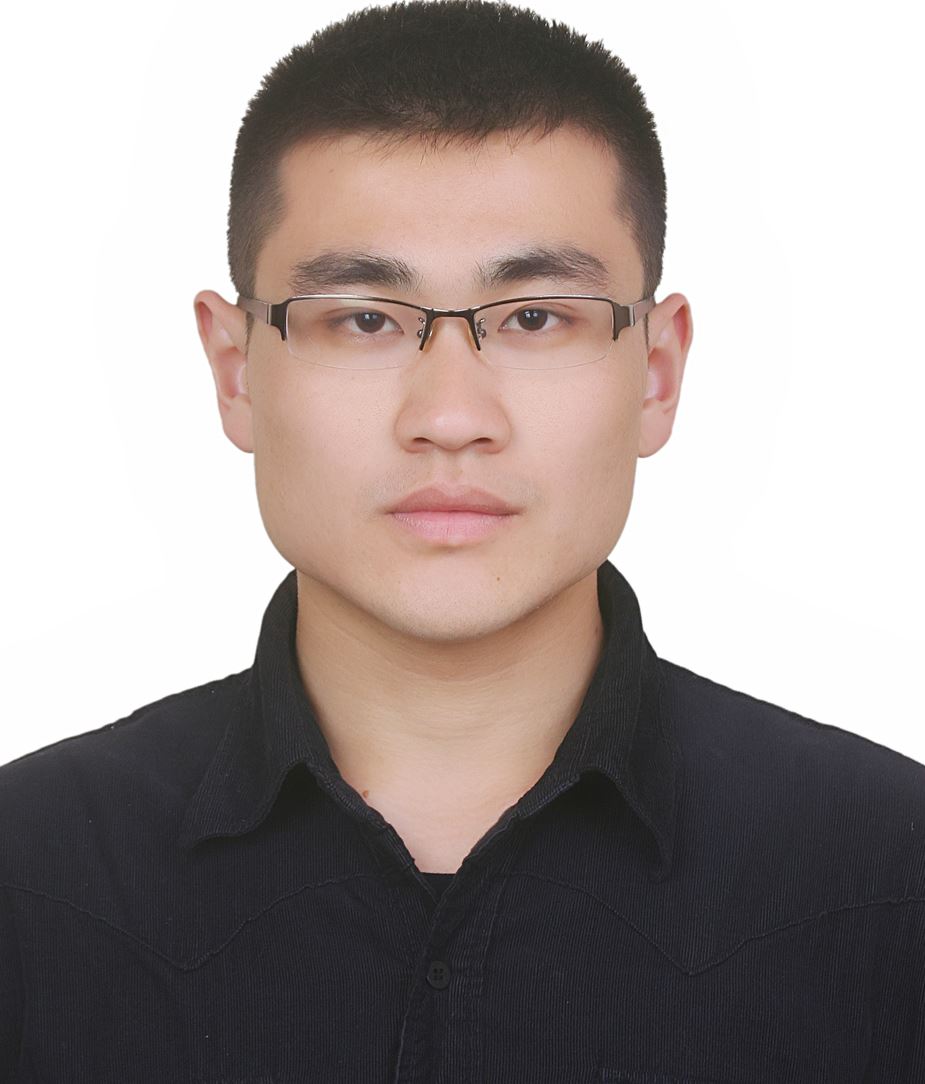}}]{Chuanyi Li}
is an Assistant Professor at Software Institute, Nanjing University, China. He received his PhD degree in 2017 from Nanjing University. His research interests include software engineering, machine learning and NLP. His research results have been published in many international journals and conferences including IEEE TSC, IEEE/ACM TASLP, ACM TKDD, JSS, Inf. Sci., ICSE, ECSE/FSE, ASE, IJCAI etc.
\end{IEEEbiography}

\begin{IEEEbiography}[{\includegraphics[width=0.9in,height=1.1in,clip,keepaspectratio]{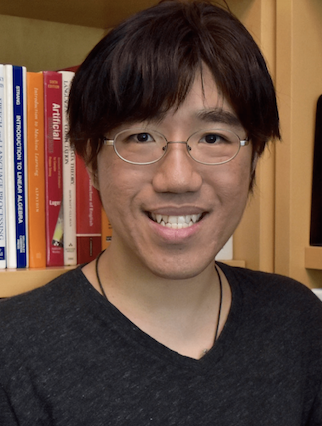}}]{Vincent Ng} got the PhD degree from the CS Department of Cornell University, in 2004. He is currently a professor and doctoral supervisor with Computer Science Department, University of Texas at Dallas. His research interests include machine learning of natural language and software engineering. His research results have been published in many international journals and conferences including IEEE TSC, ACL, ICSE, ECSE/FSE, ASE, IJCAI etc.
\end{IEEEbiography}

\begin{IEEEbiography}[{\includegraphics[width=0.9in,height=1.1in,clip,keepaspectratio]{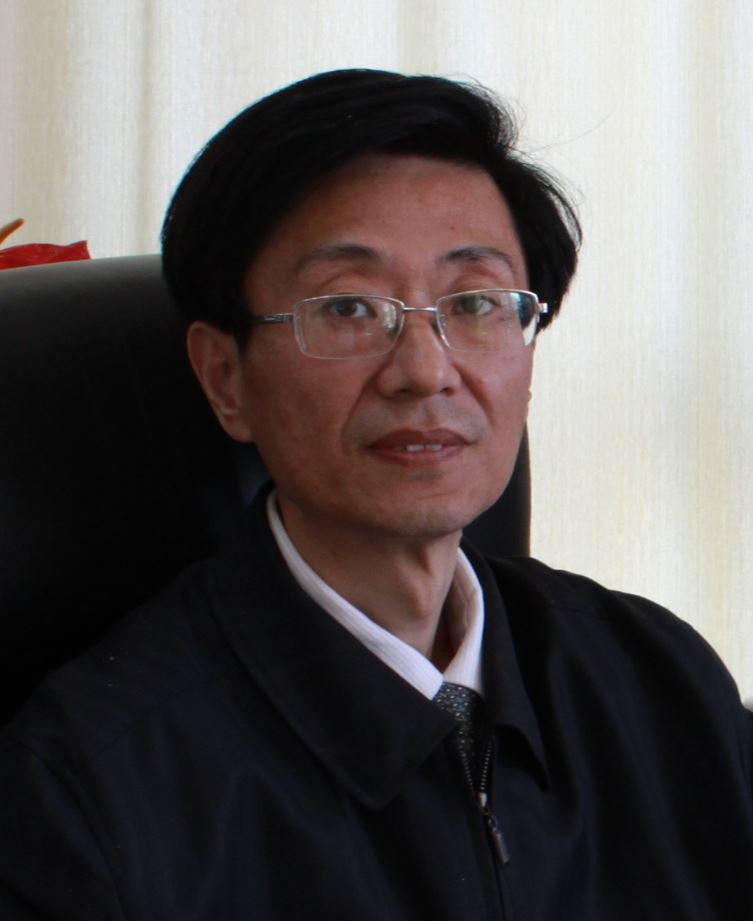}}]{Bin Luo}
is a full Professor at the Software Institute, Nanjing University. His main research interests include cloud computing, computer network, workflow scheduling, and software engineering.  His research results have been published in many international journals and conferences including IEEE TSC, IEEE/ACM TASLP, ACM TKDD, JSS, Inf. Sci., ICSE, ECSE/FSE, ASE, IJCAI etc.  
\end{IEEEbiography}

\end{document}